\documentclass[traditabstract]{aa}
\usepackage{txfonts} 
\usepackage{graphicx}

\usepackage{txfonts}

\def\la{\raise.5ex\hbox{$<$}\kern-.8em\lower 1mm\hbox{$\sim$}}
\def\ma{\raise.5ex\hbox{$>$}\kern-.8em\lower 1mm\hbox{$\sim$}}

\def\msol{M$_{\odot}$ }

\def\kms{$\rm km\, s^{-1}$}
\def\cm3{$\rm cm^{-3}$}
\def\Ts{$\rm T_{*}$~}
\def\Vs{$V_{\rm s}$~}
\def\n0{$n_{\rm 0}$}
\def\B0{$B_{0}$}

\def\Te{$\rm T_{e}$~}
\def\Tgr{$\rm T_{gr}$~}

\def\erg{$\rm erg\, cm^{-2}\, s^{-1}$}
\def\mum{$\mu$m~}

\def\L12{L$_{12\mu m}$~}
\def\F12{F$_{12\mu m}$~}
\def\agr{a$_{gr}$~}
\def\Hb{H$\beta$~}
\def\Ha{H$\alpha$~}
\def\Hg{H$\gamma$}
\def\Haa{H${\alpha}_{calc}$~}

\def\ff{{\it ff}}

\def\La{L$_{H\alpha}$~}

\def\ff{{\it ff}}

\def\RO3{R$_{[OIII]}$}

\def\Haa{H${\alpha}_{calc}$~}
\def\Hab{H${\alpha}_{obs}$~}

\def\La{L$_{H\alpha}$~}
\def\LOII{L$_{[OII]}$~}
\def\Mo{M$_{\odot}$}

\begin{document}
  \title{Analysis of the spectra observed from GRB061007, GRB061121, GRB080605, GRB090926B, GRB080207 and GRB070521
   host galaxies. \Ha and SFR trends.}

   \author{M. Contini \inst{1,2}
}

   \institute{Dipartimento di Fisica e Astronomia 'G. Galilei', University of Padova, 
              Vicolo dell'Osservatorio 3, I-35122 Padova, Italy 
                \and  
             School of Physics and Astronomy, Tel-Aviv University,
              Tel-Aviv 69978, Israel\\ 
             } 

   \date{Received }

\abstract 
{We calculate  the physical conditions and the N/H and O/H  relative abundances  
 for  a sample of long GRB (LGRB)  host galaxies 
 in the  redshift range 1$<$z$<$2.1  by modelling recently observed line and continuum spectra. 
 The  results are  consistent with those previously
calculated for LGRB host galaxies throughout a more extended redshift range z$\leq 3$. 
 We  analyse   star formation rates (SFR) within  the LGRB hosts on the basis of  the \Ha fluxes.  
 They  are compared with those of  local low-luminosity starburst (SB) galaxies, individual HII regions 
 in local galaxies as well as  LGRB host galaxies at intermediate and relatively high redshifts.
The enhanced SFR in the  HII regions within nearby galaxies is explained by  a
relatively high filling factor which  characterizes  "individual regions" rather than  "entire galaxies"  
which are generally  presented  by the observations. The  fragmented matter in the galaxies  derives 
 from  progenitor  merging.
We check whether the   release by  the morphological transformations of ice   
 of the O$_2$ and N$_2$ molecules  trapped into the ice mantles of dust grains 
 could explain the N/O ratios throughout the redshift.  
 We have found that  shock velocities  calculated by modelling the spectra are high enough to
completely destroy the ice mantles.
 Therefore, the  prevention of secondary nitrogen formation  is a valid hypothesis to explain 
 the low N/O ratios  at z$<$1.  
 The SFR trend increasing with z is roughly similar to that of  N/O.
}

\keywords
{radiation mechanisms: general --- shock waves --- ISM: abundances --- galaxies:  GRB  --- galaxies: high redshift}

\titlerunning{LGRB host galaxies at 1$<$z$<$2}
\authorrunning{M. Contini}

\maketitle

\section{Introduction}

Long GRB (LGRB) derive from implosion of massive stars. They have observed periods $>$ 2 seconds
and they are  mainly detected at redshifts z$\geq$0.08.
These data  concern the burst itself. An effort is recently  carried on  to investigate the burst influence 
on the properties of the host galaxies (e.g. Contini 2019a), for instance, 
following the shocks created by the burst as they reach and propagate  throughout the host galaxy.
 The effect of radiation accompanying  the GRB will be better
 recognized when  the  high ionization level lines emitted from the host gas will be available
 from the observations.
Meanwhile, the radiation source of the host gas is attributed to the starburst (SB) within the host. 
In a few cases (e.g. LGRB031203, Margutti et al. 2007) the spectra at different epochs were observed. 
This  allowed (Contini 2019a) to give a hint  on whether and how the host gas properties are affected by the GRB.
However, the calculated physical conditions and element abundances of all types of galaxies at 
relatively high redshifts represent only an average because the observations  generally  cover  
the  entire galaxy.
Specific observations  of different regions within the same galaxy (e.g. Nicuesa Guelbenzu et al 2015) 
will resolve this issue (Contini 2019b).

GRB host galaxies are investigated  in particular  to obtain some information about the progenitors. 
Line spectra in the optical- near infrared range are now available for most of the  survey galaxies 
also at redshifts $\geq$ 1.
The  modelling of the spectra focus  on  metallicities in order to disclose the nature
of star  forming processes throughout the redshift. It has been suggested that metallicities in GRB hosts are 
lower than solar (Asplund et al 2009)  indicating that  primordial material  
is trapped inside the emitting clouds. 
However O/H ratios higher than solar were  derived in a number of objects (e.g. Perley 2016b).
Analyzing a relatively large sample of host galaxies  of different types,
it was found (Contini 2016 and references therein)
that gas temperatures and densities  in  the emitting clouds   within GRB hosts
are  similar to those of supernova (SN) hosts and SB galaxies.
Surveys of spectral observations  including a large number of objects were presented
by Kr\"{u}hler et al (2011), Savaglio et al (2012), Han et al (2010), etc. Their  data
allowed to  calculate the element abundances  by  modelling  the line ratios.

To characterise the population of LGRB host galaxies at 1$<$z$<$3 
the conditions in which LGRBs form were addressed in the frame of 
star formation ages throughout the redshift by  
Palmerio et al (2019), Vergani et al. (2017), Hashimoto et al (2019), etc.
in order to trace star formation.
 They reported  LGRB  characteristics calculated  on the basis of large surveys of  host galaxy 
 investigations (Vergani et al 2015, Perley et al 2016a, 2016b, Japelj et al 2016a, 2016b).  
 Palmerio et al  confirmed
that LGRBs occur mainly in low metallicity  neighbourhood. However, high metallicity progenitors  were
not excluded (Berg  et al 2012, Savaglio et al 2012, Contini 2019b, etc).

Star formation rates (SFR) are generally  evaluated  by the Kennicutt et al (1998) equation on the basis of 
the \Ha observed luminosity. \Ha  are the strongest recombination lines.
However, a leading question  involving star formation  and SFR within the host galaxies  concerns
dust formation. 
Formation and destruction of stars  are closely connected with the production of interstellar dust 
(Dwek 1998, Franceschini 2001, Kreckel et al 2013)
 even in the early universe  because dust is created by the ejecta of population III stars (Nozawa et al 2003).
It was suggested  that most of the stars  at redshift z $\sim$ 1-3  were formed  in a  dusty environment 
(Le Floc'h et al. 2005, Magnelli et al. 2009, Elbaz et al. 2011, Murphy et al. 2011, Reddy et al. 2012)
implying that the infrared wavelengths are  the SFR indicators.
The UV and optical wavelengths may be  preferred as SFR indicators at very high redshift, when galaxies 
contained little dust (e.g., Wilkins et al. 2011, Walter et al. 2012) 
which could not seriously modify the
line ratios  from  their theoretical   values in  the different wavelength domains.
Calzetti (2013) claimed that the calibration of SFR indicators remains, however,  problematic for 
distant galaxies (e.g., Reddy et al. 2012; Lee et al. 2010; Wuyts et al. 2011), 
since it can be affected in particular by differences in star formation histories, metal abundances, 
content and distribution of stellar populations and dust between low and high redshift galaxies 
(Elbaz et al. 2011).  

We investigate  LGRB host galaxies focusing on the element abundances.
At present, we can determine N/H and O/H  with a satisfactory accuracy by the detailed modelling of  the
observed line ratios. When the lines relative to C, Si, S and Fe will be available from the observations
in the UV and/or in the IR range we will be able to discuss the presence of dust grains with  more
precision.  
In this paper we  start  by  modelling  the line and continuum spectra observed by Palmerio et al (2019)
and  Hashimoto et al (2019) from  LGRB hosts at 1$<z<2.1$  
 in order to add  more data to  the sample of LGRB host galaxies presented by Contini (2017 and references therein).
The choice of the survey objects is, however, limited  to  those showing   
the characteristic spectral lines which constrain the models. We also consider galaxies at lower redshifts
 which  were analysed previously  but   show new data of lines and/or of the continuum SED.
Our method  consists in calculating by the detailed modelling of the spectra 
the physical and chemical parameters 
characteristic of the gas within the galaxy which lead to the  best fit the line ratios.  
Thus, we obtain the calculated  \Hb and \Ha  fluxes 
for each galaxy.  They  can be compared with the observed ones and can be used  to calculate   
the reddening correction factor for each line ratio. 
Then,  we   analyse  SFR in LGRB host galaxies focusing on the observed \Ha fluxes. 
Comparison of calculated with observed \Ha can give a hint about SFR evolution throughout the 0$<$z$<$3 redshift.
We use composite models which account consistently for photoionization and shocks.  
The calculation code is  briefly described in Sect. 2.
In Sect. 3 the modelling of  Palmerio et al and  Hashimoto et al  observations are presented.
The distribution of the SFRs  throughout the redshift is  discussed in Sect. 4. 
Concluding remarks follow in Sect. 5.

\section{Calculation details}

The code {\sc suma} is adopted (for a more detailed  description of the code see Contini 2019a).
The main input parameters are those which lead  to the calculations of  line and continuum fluxes.
They account for photoionization and heating by  primary and secondary  radiation and for
collisional process due to shocks.
The input parameters such as  the shock velocity \Vs, the atomic
preshock density \n0 and the preshock
magnetic field \B0 (for  all models \B0=10$^{-4}$Gauss is adopted)
define the hydrodynamical field.
They  are combined in the compression equation (Cox 1972) which is resolved
throughout each slab of the gas
in order to obtain the density profile  throughout the emitting clouds.
Primary radiation for SB in the GRB host galaxies is approximated by a black-body (bb).
  The input parameters that represent the primary radiation from the SB are the 
 effective temperature  \Ts and the ionization parameter $U$. 
The primary radiation source
 does not depend on the host physical condition but it affects the surrounding gas.   This  region  is not considered
as a unique cloud, but as a  sequence of plane-parallel slabs (up to 300) with different geometrical thickness 
calculated automatically following the temperature gradient. 
The secondary diffuse radiation is emitted from the slabs of
gas heated  by the radiation flux reaching the gas and by the shock.
Primary and secondary radiation are calculated by radiation transfer.
The calculations initiate at the shock front where the gas is compressed and  adiabatically thermalised, 
reaching a maximum temperature in the immediate post-shock region T$\sim$ 1.5$\times 10^5$ (\Vs/100 \kms)$^2$. 
T decreases downstream  leading to recombination.  The cooling rate is calculated in each slab. 
The line and continuum emitting  regions throughout the galaxy cover  an ensemble of fragmented clouds.
The geometrical thickness $D$ of the clouds is  an input parameter  which is  calculated
consistently with the physical conditions and element abundances of the emitting gas.
The fractional abundances of the ions are calculated resolving the ionization equations
for each element (H, He, C, N, O, Ne, Mg, Si, S, Ar, Cl, Fe) in each ionization level.
Then, the calculated line ratios, integrated throughout the cloud geometrical width, are compared with the
observed ones. The calculation process is repeated
changing  the input parameters until the observed data are reproduced by the model results,  at maximum
within 10-20 percent
for the strongest line ratios and within 50 percent for the weakest ones.

The gas ionized by the  SB  radiation flux  emits continuum radiation (as well as the line fluxes)
from  radio to  X-ray.
The continuum accounts for free-free and free-bound radiation (hereafter  addressed to as bremsstrahlung).
Some parameters regarding  directly the continuum SED, such as the dust-to-gas  ratio $d/g$  and the dust 
grain radius \agr ~ are not directly constrained by fitting the line ratios.
Dust grains are sputtered throughout the shock front and downstream. They are  heated by the primary radiation 
and by mutual collision with  atoms.
The dust reprocessed radiation in the infrared (IR) range throughout the SED   depends on $d/g$ and \agr.
The distribution of the grain sizes along the cloud starting from an initial radius
is automatically  calculated  by {\sc suma}.

\section{Modelling the spectra}

\subsection{Line ratios}

\begin{table*}
\centering
\caption{Observations and models}
\begin{tabular}{lccccccccccccc} \hline  \hline
\ GRB    & z         & [OII]3727+ & [NeIII]3868 & \Hg   & \Hb  & [OIII]5007+ &  \Ha        & [NII]6583 \\ \hline
\ 061007$^1$ &1.2623 &2.4         &$<$2         & -     &  1.0 &9.5+         &4$\pm$0.4    & $<$2.4    \\
\ corr   &           &2.98        &$<$2.42      & -     & 1    &8.7+(2.2)    &3            & $<$1.8    \\
\ mod1   &           & 3.3        & 1.0         & 0.46  & 1    & 11.74       &3            & 0.36       \\
\ 061121$^1$ &1.3160 &3.38        &0.32         &0.53   &1     &4.36         &5.06$\pm$0.9 & 0.57       \\
\ corr   &           &4.97        &0.46         &0.64   &1     &3.4          &3            & 0.34       \\
\ mod2   &           &5.2         &0.55         &0.46   & 1    &3.47         &2.96         & 0.39       \\
\ 080605$^1$ &1.6408 &2.22        & -           & -     & 1    &5.18         &3.78         & 0.52       \\
\ corr   &           &2.64        & -           & -     & 1    &5.00         &3            & 0.4        \\
\ mod3   &           &2.5         & -           & -     & 1    &5.02         &2.94         & 0.33       \\
\ 090926B$^1$&1.2427 &4.96        &$<$0.92      &$<$1.17& 1    &6.38         &4.79         & $<$0.8     \\
\ corr   &           &7.07        &$<$1.13      &$<$1.38& 1    &5.8          &3            & $<$0.8     \\
\ mod4   &           &7.1         & 0.75        &0.46   & 1    &5.9          &2.95         & 0.5        \\
\ 080207$^2$ & 2.086 & 2.028      & 0.73        &-      &1     & 4.836       & 3.0         & 0.68        \\
\ mod5   &           &1.8         &0.5          &0.46   &1     &4.9          &3.2          &0.52       \\ \hline

\end{tabular}

	$^1$ from Palmerio et al (2019); $^2$ from Kr\"{u}hler et al (2015)

\end{table*}

\begin{table*}
\centering
\caption{Model results}
\begin{tabular}{lccccccccccccc} \hline  \hline
\  models            &mod1   & mod2   & mod3   & mod4     & mod5   \\ \hline
\ \Hb$^1$            & 0.02  & 0.0023  & 0.03  & 0.0094   &  0.084   \\
\  \Vs (\kms)        &200    & 100     & 200   & 160      &320 \\
\ \n0  (\n0)         &60     & 90      &80     & 110      &80 \\
\ $D$  (pc)         &0.23   & 0.02    &0.1    & 0.1     & 1.1\\
\ \Ts  (10$^4$K)    & 8.8   & 7.2     &5.4    & 9.4   &9.7   \\
\ $U$ -              &0.04   & 0.0034  &0.042  &0.0085 &0.04\\
\ He/H               &0.1    & 0.1     &0.1    &0.1    &0.1\\
\ N/H$^2$            &0.22   & 0.2     &0.4    &0.2    &0.5\\
\ O/H$^2$            &6.6    & 6.6     &6.6    &6.6    &6.6\\
\ Ne/H$^2$           &1.     & 1.      &1.     &0.7    &1  \\
\ S/H$^2$            &0.2    & 0.3     &0.3    &0.2    &0.2\\
\ Ar/H$^2$           &0.06   & 0.04    &0.03   & 0.006 &0.06\\   
\ 12+log(O/H)$^3$   &8.82   & 8.82    &8.82   &8.82   &8.82\\
\ 12+log(O/H)$^4$   &8.13   &8.51      &8.47  &8.48   & 8.74$^5$\\ \hline

\end{tabular}

$^1$ \Hb absolute flux calculated at the nebula in \erg ;
	$^2$ in 10$^{-4}$ units; $^3$ calculated in this paper; $^4$ Palmerio et al;
	$^5$ calculated by Kr\"{u}hler et al (2015).

\end{table*}

\begin{figure*}
\centering
\includegraphics[width=7.8cm]{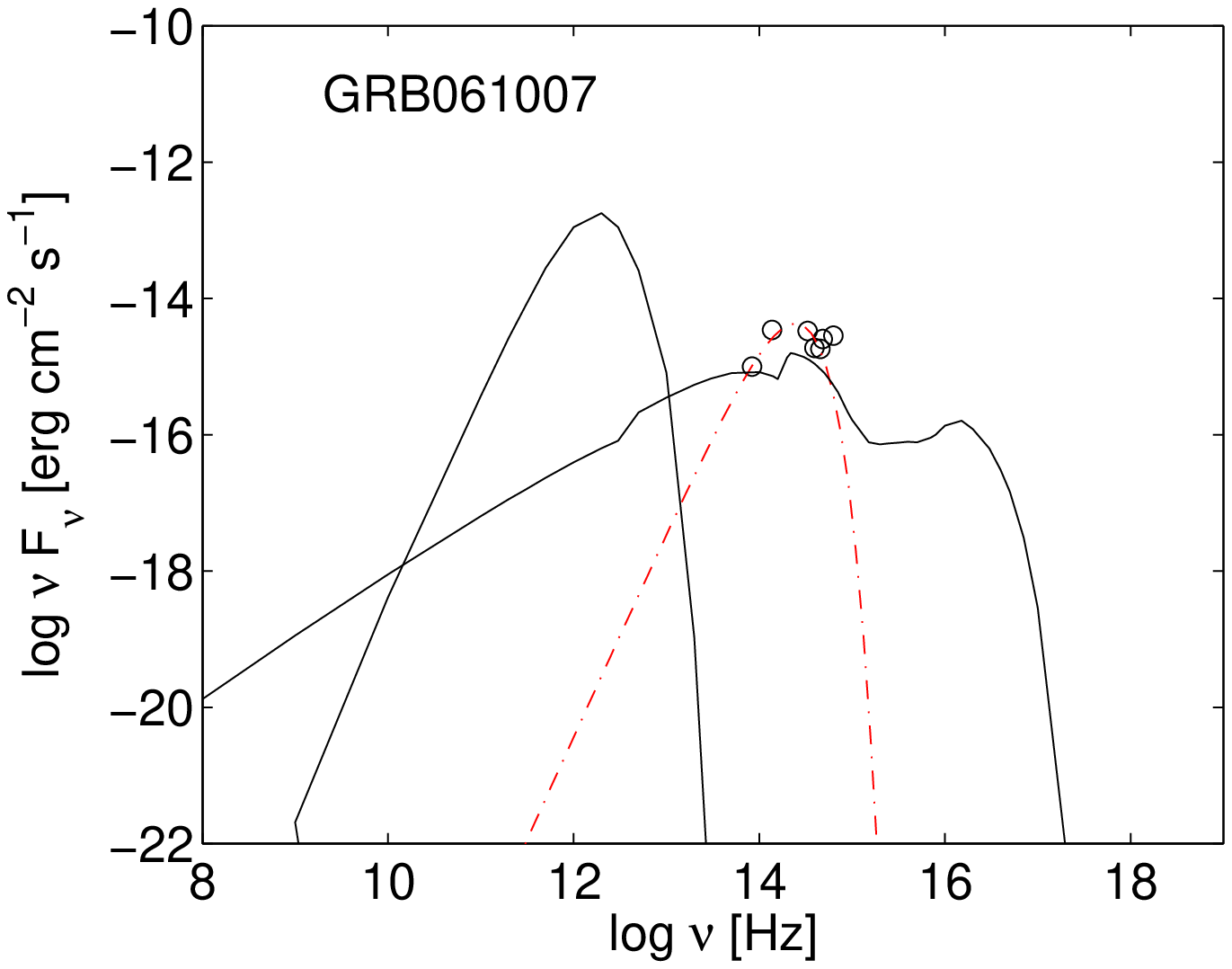}
\includegraphics[width=7.8cm]{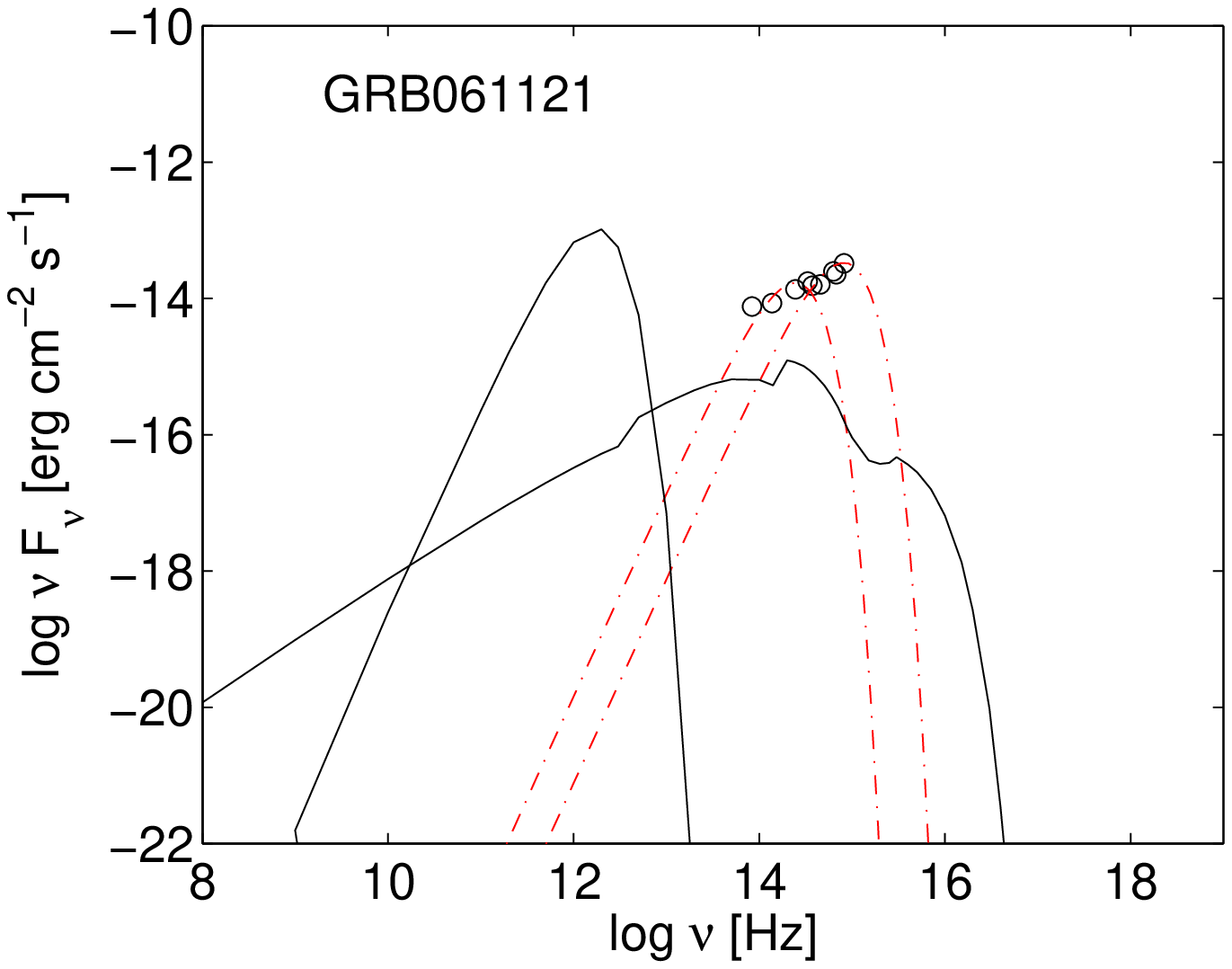}
\includegraphics[width=7.8cm]{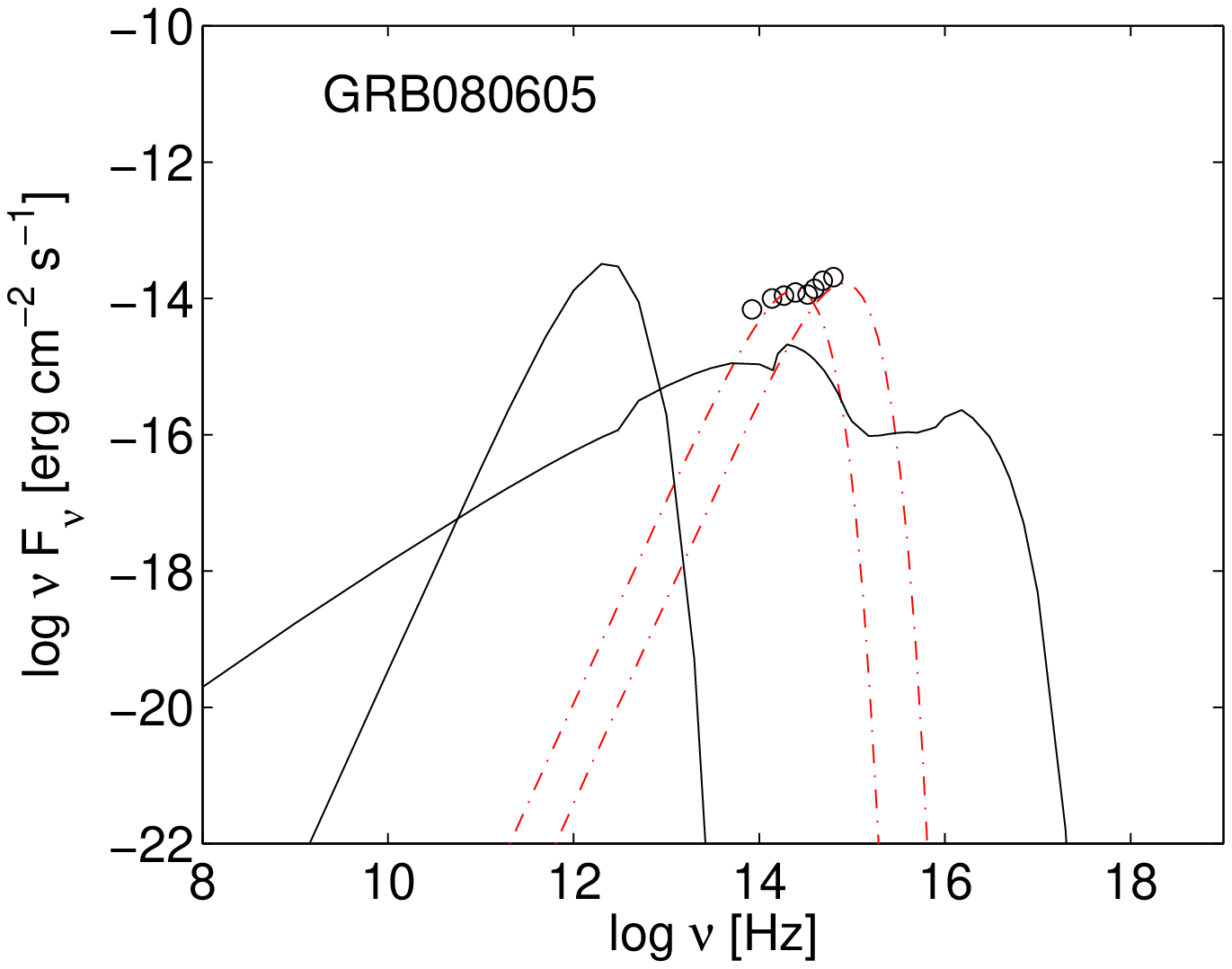}
\includegraphics[width=7.8cm]{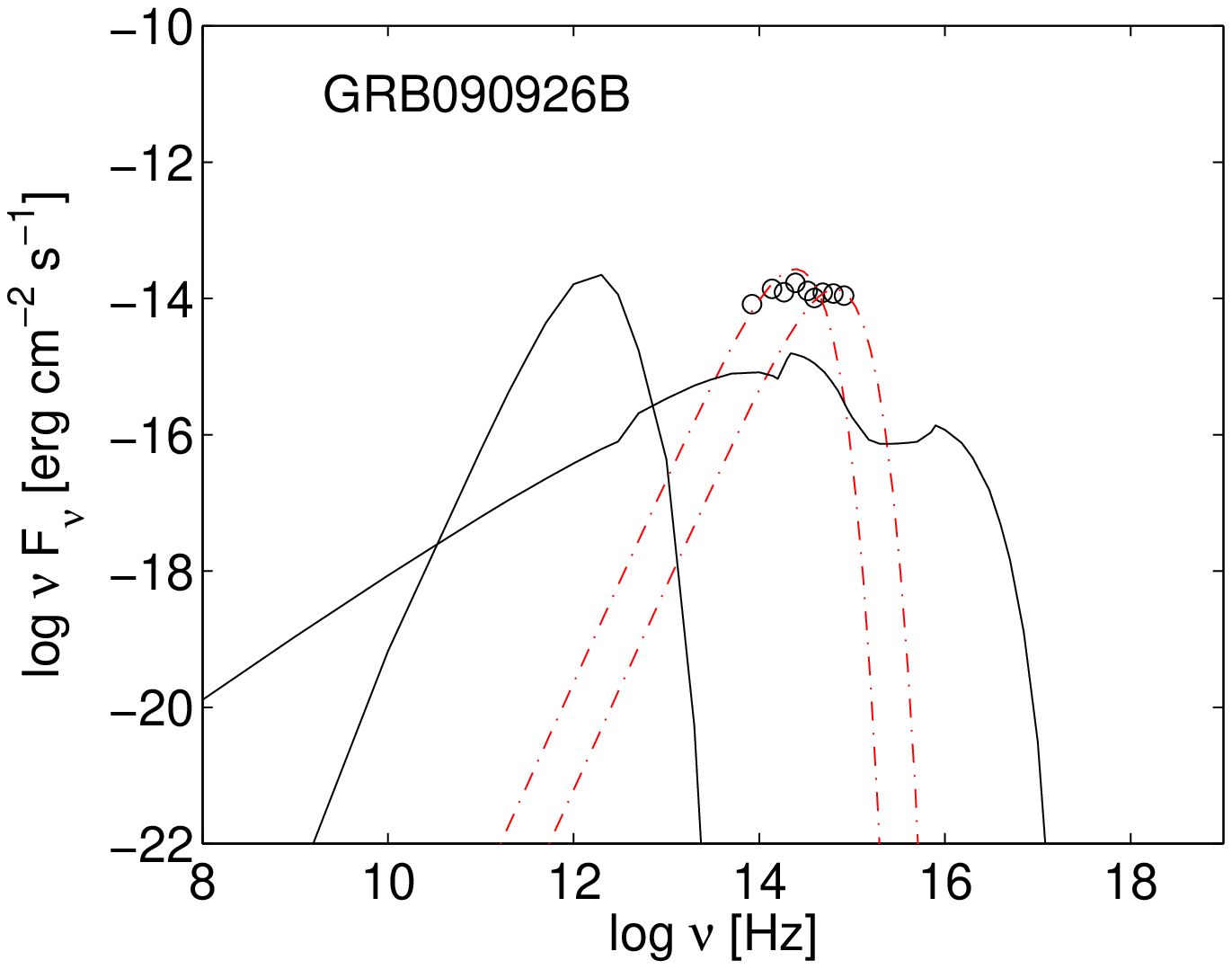}
\caption{SED of GRB061007, GRB061121, GRB080605 and GRB090926B. Open circles : the data from Palmerio et al.
Red dot-dashed line : the contribution from the underlying old stars. Black solid lines: the bremsstrahlung
corresponding to models (mod1 - mod4) and to reprocessed radiation by dust in the
IR calculated consistently by the models.
}
\end{figure*}

\begin{figure}
\centering
        \includegraphics[width=9.2cm]{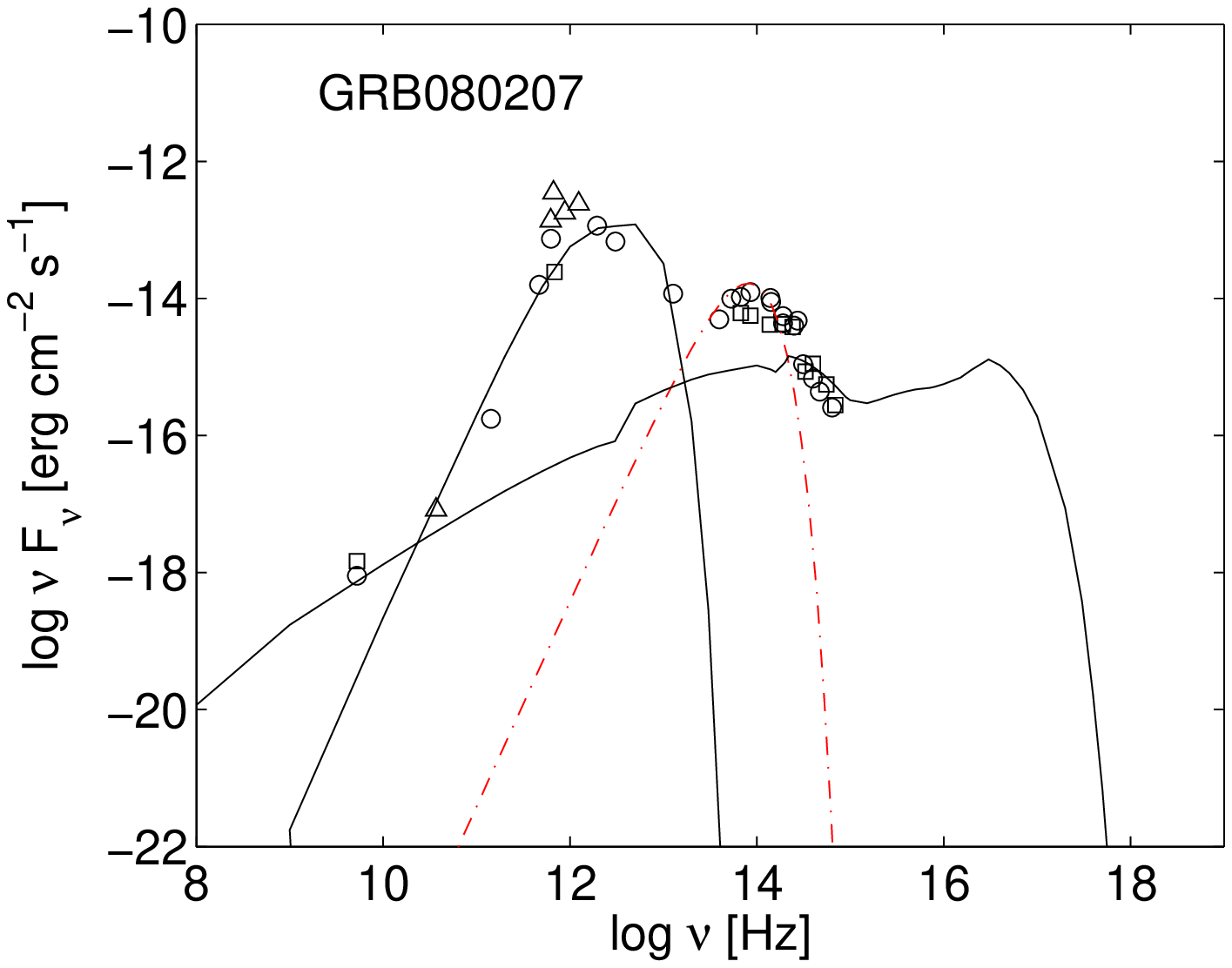}
\caption{
        SED of GRB080207. Black open circles: data from Hashimoto et al (2019).
	Open squares : data for GRB070521.
        Black open triangles : upper limits. Solid black line : bremsstrahlung and dust
	reprocessed radiation calculated by fitting  Kr\"{u}hler  et al (2011) data
by Contini (2016).
Red dash-dotted line: black-body flux at 1000K from the underlying stellar population.
}
\end{figure}

Palmerio et al (2019)  report the sample of LGRB host galaxies at  1$<$z$<$2 (Salvaterra 2012) 
observed by  the  Swift/BAT6 (Burst Alert Telescope).
We  consider for modelling   Palmerio et al   data   presented in their table A.4. 
The observed line fluxes are given in units of
10$^{-17}$ \erg and  are corrected for Galactic foreground extinction.
The modelling procedure requires a minimum number of specific lines in order to constrain the models.
We select the galaxies which show  \Ha, \Hb and   oxygen   in at least two  ionization levels,
GRB061007, GRB061121, GRB08605 and GRB090926B.
They  are reported in Table 1. For each galaxy the  row showing the observed flux ratios  is followed  by
the line intensity ratios to \Hb  (reddening corrected).
We  have further corrected the line ratios using Osterbrock (1974) equations  when \Ha/\Hb $\geq$ 4  because
 the  \Ha/\Hb  ratios calculated by quantum mechanic procedures  range
between  3.05 and 2.87   for gas temperatures  between $\sim$ 5000 K and 10000 K and
 preshock densities   \n0 $\sim$  10$^2$-10$^4$ \cm3.
These values are similar to those calculated by the detailed modelling of the spectra 
(which means that all the lines are considered) for most of the LGBR host galaxies. 
 The preshock densities calculated for the galaxies in Table 1 are not as high as to yield \Ha/\Hb $>$4.
These values can be found for  high optical depths (Osterbrock 1974, fig 4.3).
In Table 1, for each host galaxy, below the row  reporting the corrected line ratios, the model calculation results are
shown for comparison.
For the host of GRB061007 the [OIII] 5007 line is given by the observations, while the [OIII] 4959 line is omitted. 
The two lines  belong to the same multiplet.  From the ratio of the 
the  transition probabilities (Osterbrock 1974) the [OIII] 4959 line should be $\sim$ 1/3 [OIII] 5007.
On this basis we have added the [OIII] 4959  line flux for GRB061007 in Table 1. 

 In the bottom of Table 1  the line ratios for  the host galaxy GRB080207 at z=2.086 are shown. 
 We  added in Table 1 the host line  spectrum  of this object  from the  sample of Kr\"{u}hler et al (2015)  because
 the model can be constrained by Hashimoto et al (2019)  rich photometric data of the continuum SED.
 Hashimoto et  al reported for GRB080207 also the [CII] 156 \mum line in the far-IR.
 Model calculations  yield [CII]/\Hb = 0.04 and \Hb= 0.09 \erg. 
For  GRB080207, after the GRB detection by SWIFT/BAT (Racusin et al 2008), 
no optical and NIR afterglows were detected. The extremely red host galaxy was identified within 
the X-ray positional
error circle (Hunt et al 2011, Svensson et al 2012). The \Ha and [OIII]5007 lines
were  observed with VLT/X-Shooter by Kr\"{u}hler et al (2012) yielding  SFR$\sim$ 77 \msol yr$^{-1}$.
The host galaxy was also detected by Herschel/PACS at $\sim$ 30\mum and 50\mum.
 The modelling of the line ratios  is reported by Contini (2016, table 8).

In Table 1 we  compare the results of model calculations  with the observations.
The calculation  uncertainty is  $\sim$10 percent.
The  models are represented by the sets of the input parameters which lead to 
the best fit of the observed line ratios. They are described in Table 2. 
We adopt for all  the spectra a  black-body dominated model representing the star burst.
The effective temperatures calculated for the star burst in the host galaxies
are high enough   to indicate  a quite recent event.
The O/H relative abundances calculated by the detailed modelling are  solar 
((O/H)$_{\odot}$=6.6 10$^{-4}$, Asplund et al 2009)  for all the hosts, 
while the N/H ratios are low relatively to the solar ones
((N/H)$_{\odot}$=10$^{-4}$) and  similar to those of other LGRB hosts.
The results suggest that  LGRB at z$<$ 2 occur in a low metallicity medium, relative to nitrogen.
In the bottom of Table 2 metallicities in terms of 12+log(O/H) calculated by the present detailed modelling 
and by the strong line method
by Palmerio et al are compared.  Palmerio et al values are lower than solar, therefore they  claim that
SFR  should be high in their sample of host galaxies on the basis of  the low metallicity. 

\subsection{Continuum SED}

The bremsstrahlung emitted from the galaxy clouds  can be seen throughout the SED in  the X-ray - radio range.
The dust reradiation bump in the infrared and the old star backgound population flux emerge  
from it in nearly all the  galaxy types.
In this work we use $d/g$=10$^{-14}$ by number for all the models which corresponds to
4.1 10$^{-4}$ by mass for silicates (Draine \& Lee 1994).
The black body emission from the background old star  population with
T$_{bb}$$\sim$ 3000-8000 K  generally covers the near-IR (NIR) - optical range of the SED.

The modelling of the continuum SED of each object is  calculated   by the same model which fits the line ratios.
The results are shown in the  diagrams  of Figs 1 and 2.
Two main solid lines  show the result  of modelling.  One represents the bremsstrahlung
emitted from the gaseous clouds within the galaxy which also emit the lines, 
while the other  line represents  reprocessed radiation by dust.
The red dash-dotted lines indicate the black-body flux emitted from the underlying old star 
population  with different temperatures. 
For the Palmerio et al sample which appears in Fig. 1 diagrams the modelling of the continuum   SED
is not constrained by the   data in the radio, in the far-IR, not   even in the very NIR, while  the model results
are definitively constrained by the data reported by Hashimoto et al  for GRB080207 (Fig. 2). 
Therefore, in Fig. 1 diagram, we have normalized the bremsstrahlung radiation
by the frequency in the radio range at $\nu$= 10$^8$ Hz, waiting for future data. 
The underlying  old stellar temperatures which result from the IR-optical bump is $\sim$ 3000 K for GRB061007,
while  the  temperatures  range between
 3$\times$10$^3$K- 10$^4$K, 3$\times$10$^3$K- 10$^4$K and 3$\times$10$^3$K- 8$\times$10$^3$K
for GRB061121, GRB080605 and GRB090926B, respectively.

The relatively  rich dataset presented for the SED by Hashimoto et al for GRB080207 (Fig. 2) covers a more 
extended frequency range,
accounting for the radio range, the IR range  of dust reprocessed radiation and the NIR-optical range
due to the underlying stellar population. 
This  host continuum SED  in the  far-IR is reproduced  by a dust-to-gas ratio by 
mass  $\sim$  4.1 10$^{-4}$. 
GRB070521   was detected by the burst alert telescope (BAT) by Gehrels (2004) and Guidorzi et al (2007a).
The X-ray afterglow was detected by the Swift X-Ray Telescope (XRT) by Guidorzi (2007b).
The redshift of the host galaxy at z=2.0865 was determined by the \Ha emission by Kr\"{u}hler et al (2015)
leading to SFR=26 \msol yr$^{-1}$. 
The line fluxes observed from GRB070521 and from other samples, as e.g. those of Vergani et al (2017) could not be used 
for modelling because  the data were not enough to constrain the models.
On the other hand, Hashimoto et al (2019) presented a rich photometric set of data also for this GRB host.
We try to  reproduce  the observation data in Fig. 2  by the same model as that  calculated for GRB080207. 
The fit is  suitable enough  to
suggest  that similar physical conditions  characterize both the GRB070521 and  GRB080207 host galaxy gas.
 The  black-body bump in the IR (red line), which represents the contribution of the underlying old star population,
 has a temperature of 1000 K
which is lower than generally observed for other GRB hosts and it  is most likely interpreted as emission 
by hot dust rather than by old stars. 
Dust grains can be  heated to T$\sim$1000 K (close to evaporation) by strong radiation and/or collisionally by  
gas  throughout 
strong shocks (\Vs$>$500 \kms) and downstream.
These conditions are  not easily found  in LGRB host clouds.
We have adopted a maximum radius
\agr =1 \mum for the dust grains  in all the galaxies presented in Figs. 1 and 2.
  Fig. 3  shows the profiles of the grain radius \agr sputtered downstream  by the shock,  of the grain temperature
\Tgr and of the gas temperature \Te throughout  the clouds in the GRB080207 host galaxy. 
Large dust  grains  with \agr$\sim$ 1 \mum survive sputtering within the clouds 
because the shock velocity (Table 2, mod5)  is  low enough ($<$350 \kms).  
For the other galaxies  \agr  remains constant throughout the clouds due to the low \Vs ($\leq$200 \kms).
The fit of the GRB080207 dust reprocessed radiation bump leads to a maximum \Tgr  $\sim$ 70 K. 
On the other hand,  dust shells with  temperatures of $\sim$ 1000 K are 
present in late-type stars (e.g. Danchi et al. 1994) as well as in symbiotic systems (Angeloni et al 2010) 
and  can explain the IR black body radiation  bump  of  the underlying stellar population.

\begin{figure}
\centering
\includegraphics[width=4.4cm]{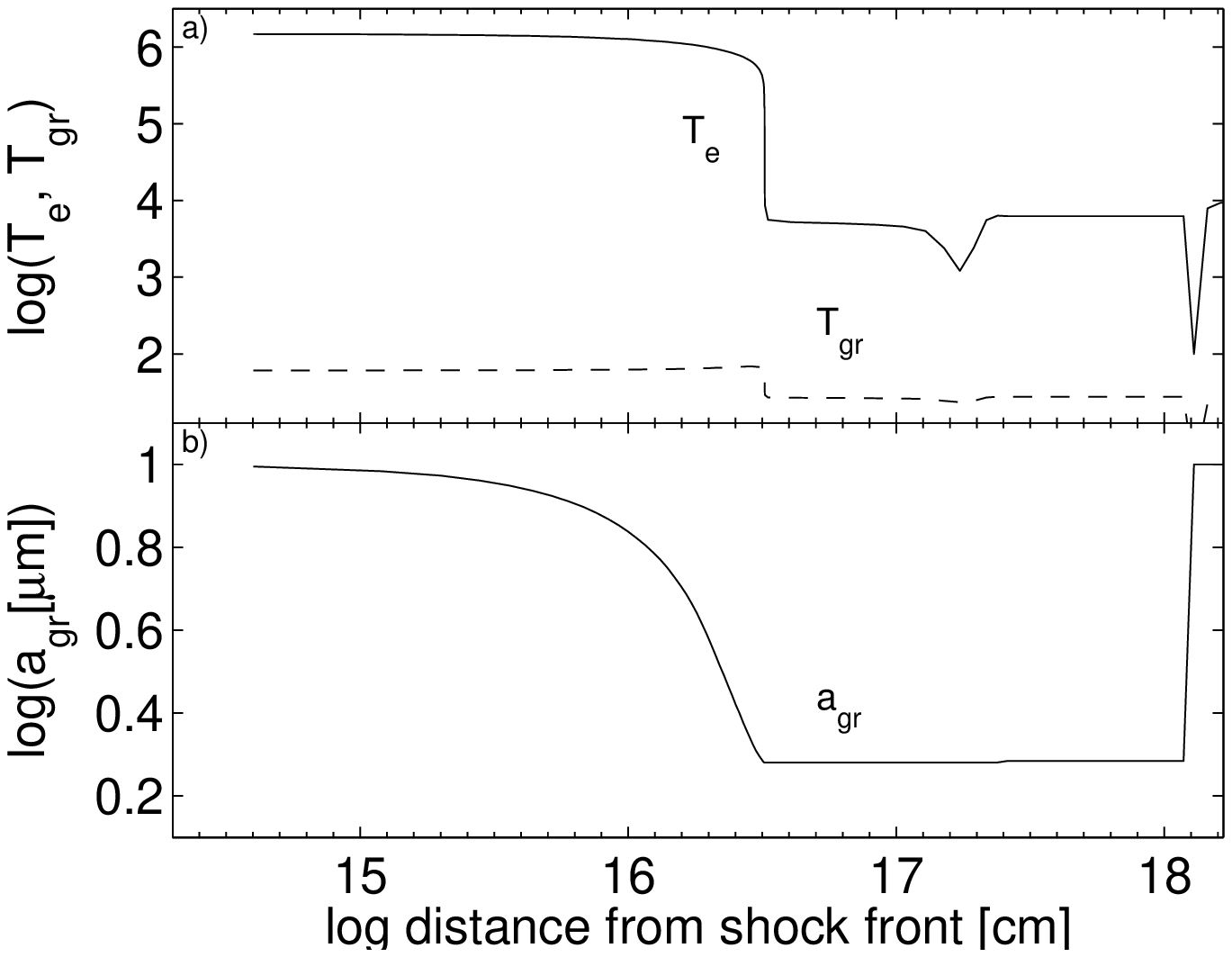}
\includegraphics[width=4.4cm]{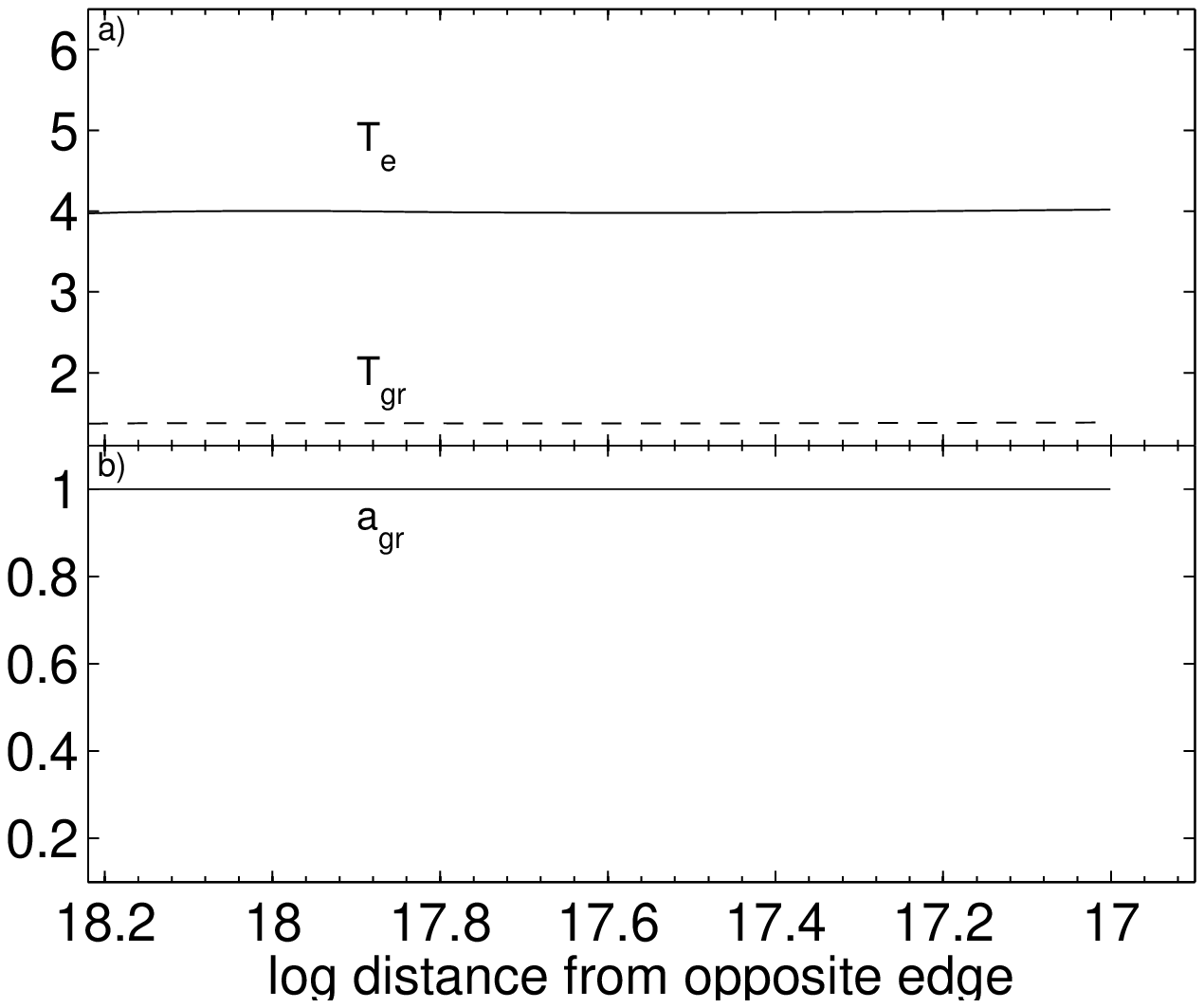}
	\caption{
		Top diagrams: the profile of the gas temperature (solid line) and grain temperature (dashed line)
		throughout a cloud in the LGRB080207 host. 
	Bottom diagrams: the  profile of the grain radius in \mum. The cloud is divided into two halves.
	The left panel shows (in logarithic scale) the region downstream of the shock front which is 
	at the left edge of the left panel.
	The right panel shows (in reverse logarithmic scale) the region of gas  reached 
	by the  radiation flux on the right edge of the right panel.}

\end{figure}

\section{SFR evolution trend}

Kennicutt et al (1998) investigating SFR  following the evolutionary properties of galaxies claim that 
most information comes from the integrated UV, far-IR (FIR) flux and from nebular
recombination lines. The last ones provide a "sensitive probe of young massive star population",
while FIR efficacy as a SFR tracer depends on the contribution of young stars to heat the dust grains.
Generally SFR are given by the observers on the basis of the \Ha line flux.
We  therefore investigate  SFR at different  redshifts by the analysis of the \Ha luminosities
adopting  Kennicutt et al (1998) equation:

\noindent
SFR (\msol y$^{-1}$) = 7.9$\times$ 10$^{-42}$ \La (erg s$^{-1}$)  (1)

\noindent
because 
SFR (\msol y$^{-1}$)= 1.4$\pm$0.4$\times$10$^{-41}$\LOII (erg s$^{-1}$) (2)

\noindent
which was written for the [OII] line (\LOII is the [OII] luminosity),  leads to  less  correct  results
(see Kennicutt et al).

We have checked the spectra  by Palmerio et al  and  Kr\"{u}hler et al in Table 1.   Not all of them show [OII]3727+/\Hb 
intensity ratios higher than the \Ha/\Hb ones. As for the sample of LGRB hosts  adopted  by Contini (2016, 2017)
the [OII]3727+/\Hb line ratios are  seldom higher than the \Ha/\Hb. 
Therefore  we will  consider eq. (1) valid for the calculation of SFRs.
\La, the \Ha luminosity observed at Earth, is the same as that calculated at the nebula i.e.
\La=\Hab$\times$4$\pi$d$^2$=\Haa$\times$4$\pi$R$^2$, where d is the distance from Earth, 
 \Hab is the \Ha flux observed at Earth,  \Haa is the \Ha flux
  calculated at the nebula and R the radius of the emitting nebula in terms of
  the distance of the nebula from the galaxy center.
Model results are  valid for the emitting nebula, while the observations 
are obtained at Earth, therefore  the \Hab/\Haa ratios range within a factor of $\sim$ 10$^{-16}$ - 10$^{-17}$, 
depending on the distance of the emitting galaxy.  R covers a range of 0.01-100 pc for LGRB host clouds.
 The observed data are averaged on the whole galaxy, therefore a filling factor
 should be accounted for   because the morphological  distribution of matter throughout the galaxies
 is highly fragmented.

\begin{figure}
	\centering
	\includegraphics[width=8.6cm]{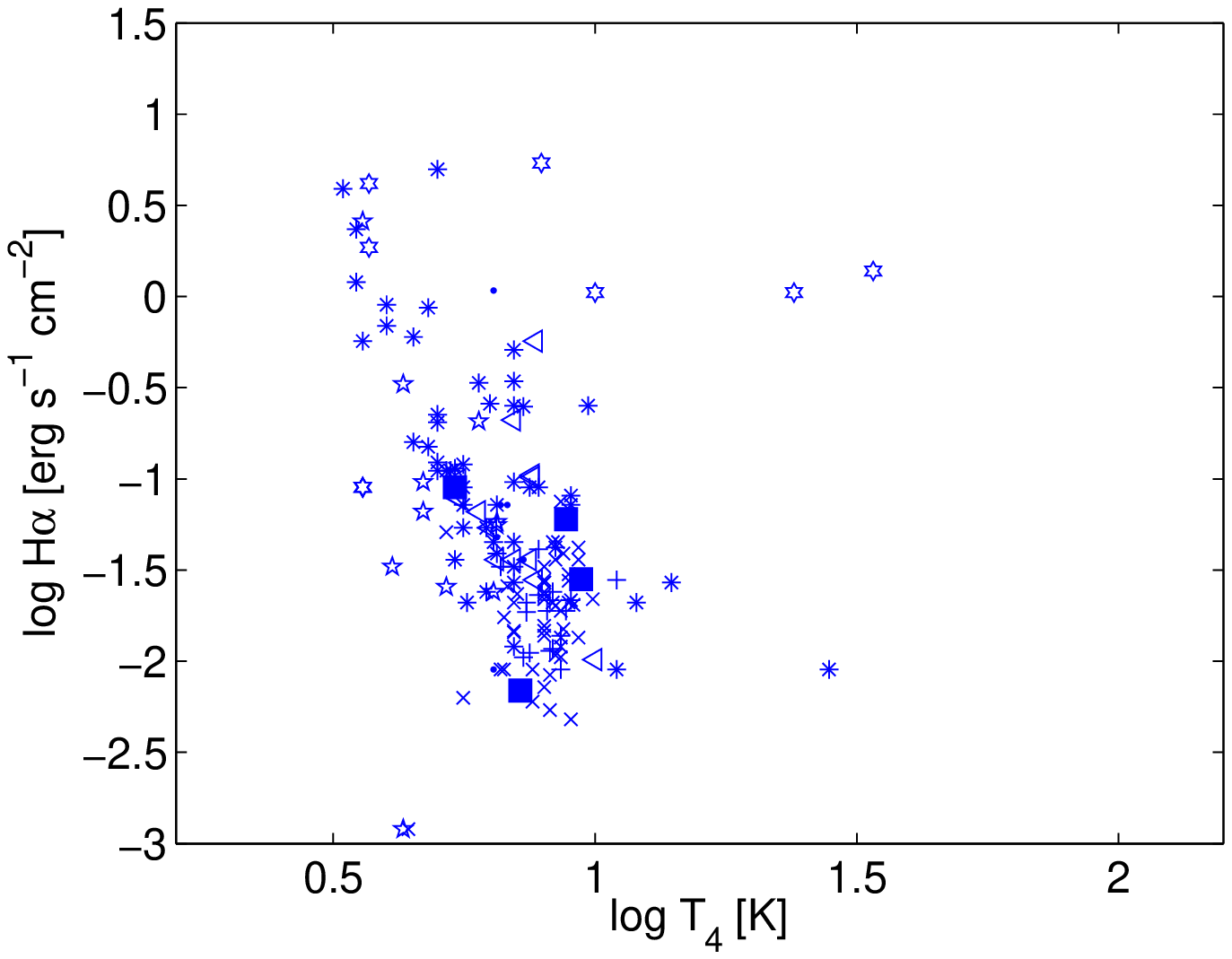}
	\includegraphics[width=8.6cm]{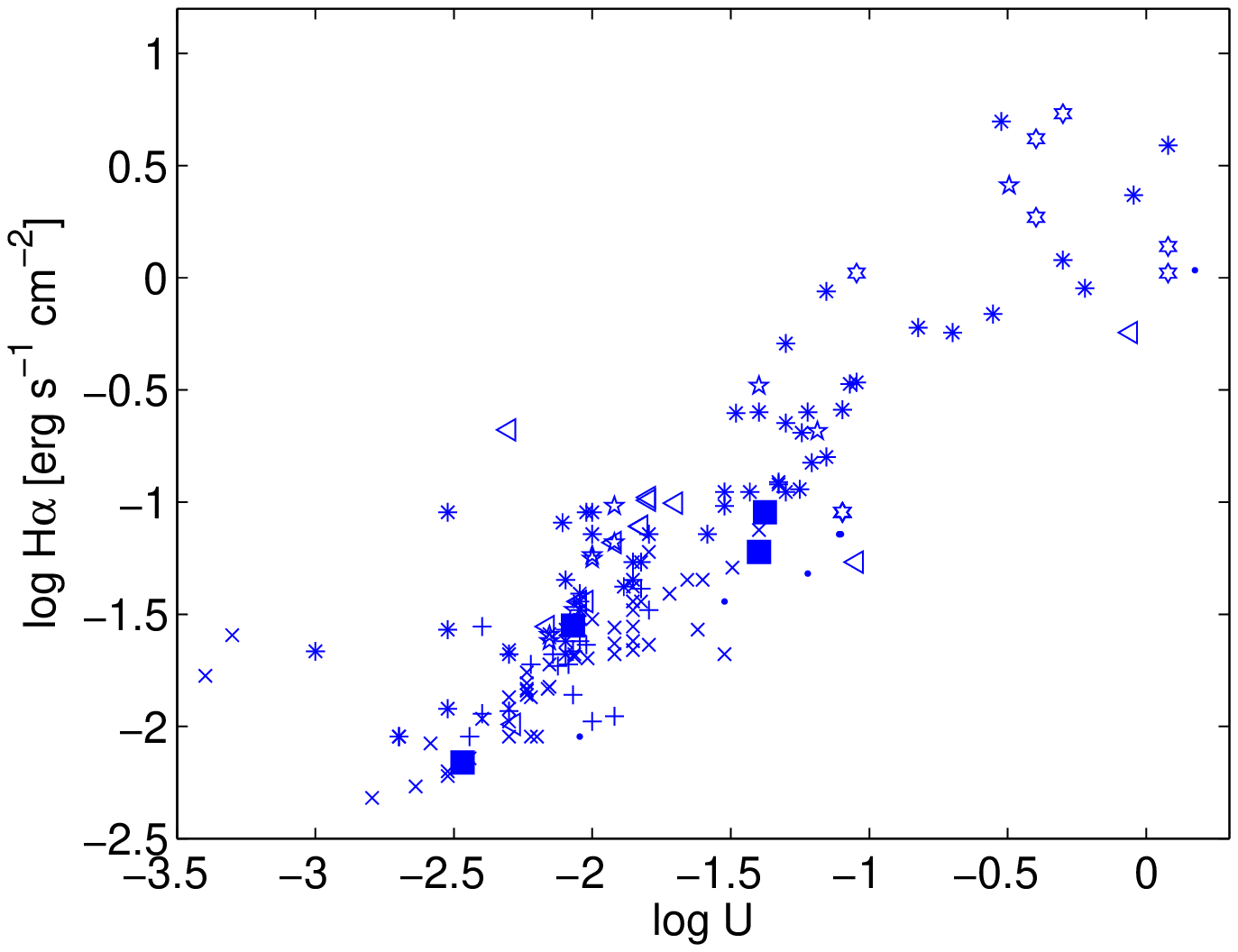}
	\includegraphics[width=8.6cm]{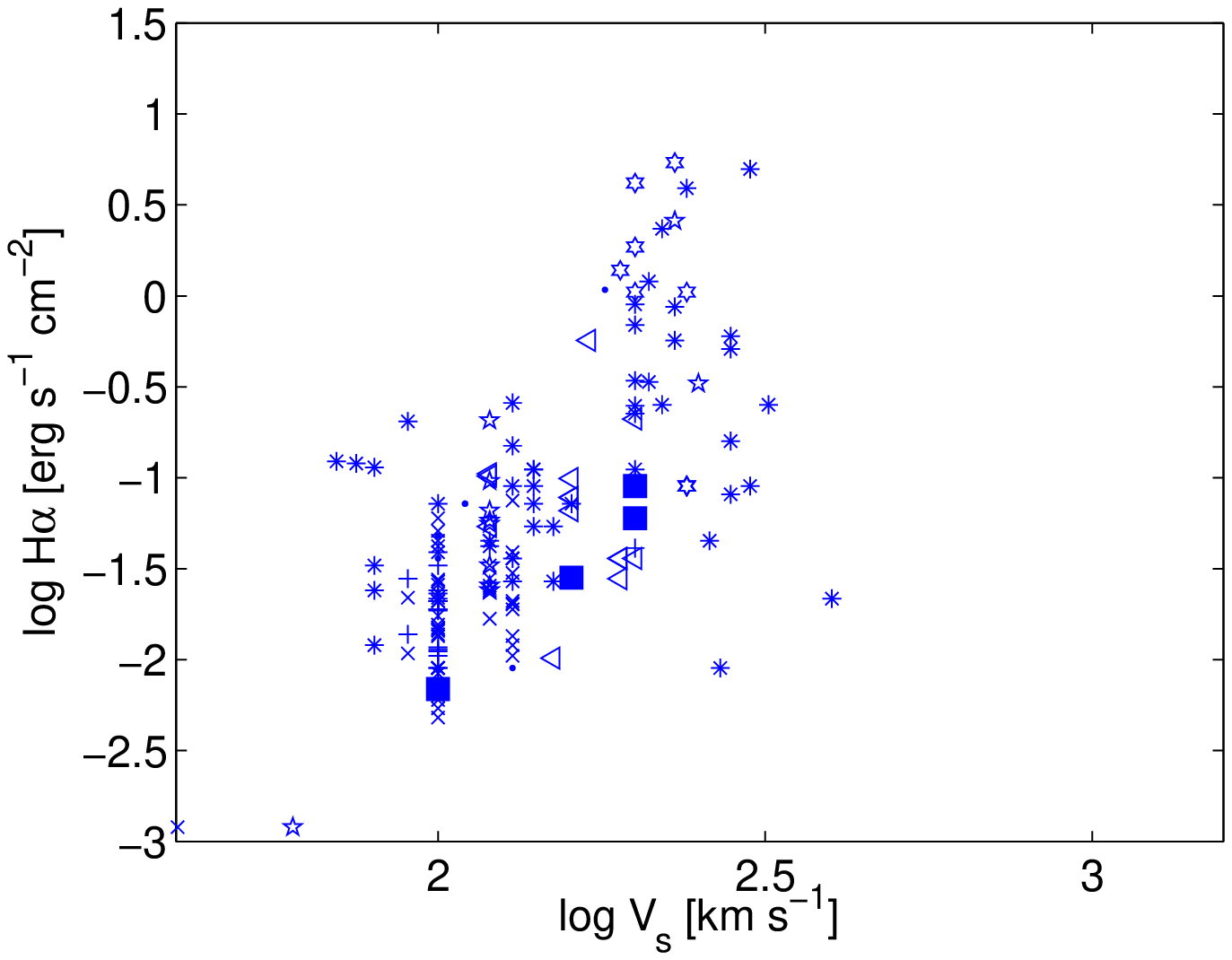}
	\caption{\Haa  as  function of \Ts (in 10$^4$K), $U$ and \Vs (in \kms).
Symbols are explained in Table 3.
}

\end{figure}

 \begin{table}
        \centering
        \caption{Symbols in Fig. 4}
        \begin{tabular}{llcl} \hline  \hline
                \ symbol   &object                       & Ref. \\ \hline
                \ asterisks &GRB hosts                   & (1) \\
                \ point      & LGRB hosts         & (2)  \\
                \ pentagrams  & LGRB different hosts     &  (3)  \\
                \ hexagram & LGRB hosts with WR stars& (4) \\
                \ triangles &LGRB at low z     & (5)\\
                \ filled squares    &LGRB in this paper& (6) \\
                \ plus&  HII regions in local galaxies   & (7)  \\
                \ cross&  HII low-luminosity nearby galaxies & (8)  \\ \hline
        \end{tabular}

        (1) (Kr\"{u}hler et al. 2015);
        (2) (Savaglio et al. 2009);
        (3) (Sollerman et al (2005), Castro-Tirado et al. (2001), Graham \& Fruechter (2013),
                Levesque et al. (2010), Vergani et al. (2011), Piranomonte et al. (2015);
                (4) (Han et al. 2010);
                (5) (Niino et al. 2016);
                (6) (Palmerio et al 2019);
                (7) (Marino et al. 2013);
                (8) (Berg   et al. 2012);

\end{table}

\subsection{\Ha line flux}

In order to understand the SFR trend  throughout a large  redshift range
we  show in Fig. 4 the
\Ha fluxes  calculated  at the nebula on the basis of the corrected observed spectra,
in order to avoid distance problems.
  In our  previous investigation  of SFR in LGRB host galaxies (Contini 2016, fig. 8) we   presented
  the  physical parameters which contribute to  \Ha. 
  Line fluxes   are calculated  by the models which lead to the best fit of the
    spectra  observed  from different objects at different redshifts.
  The line intensities  result from  summing the contribution of the different slabs of clouds within the gas
nebula.
The effective Balmer line recombination coefficient $\alpha_{eff}$ is $\propto$ 2$\times$10$^4$/T$_e^{0.9}$
  (Brocklehurst 1971). 
    The calculation of the \Ha flux accounts also for the most  significant parameters (e.g. \Ts, $U$, \Vs, \n0 and $D$).
We  investigate the trends of \Ha calculated at the emitting nebula (\Haa)
as function of the different physical parameters. 
They   are  shown in Fig. 4  on the basis of  the  diagrams previously presented by Contini (2016, fig. 8).  
We have added \Haa calculated by  the data  
given  by Berg et al  (2012) for low luminosity local HII galaxies and
 by  Marino et al (2013) for the Calar Alto Legacy Integral Field Area (CALIFA) sample
 of individual HII regions in nearby galaxies 
 in order to  enlarge the redshift range by  including  local galaxies.
Also, the  results of modelling  the  LGRB host spectra at intermediate and low z (z$\leq$0.4) by
Niino et al (2016) are shown in Fig. 4 diagrams as well as
 \Haa  calculated by modelling the data  presented by
 Palmerio et al (2019)  in the 1$<$z$<$2.1 range.
The diagrams   showing  a  clear trend of \Haa with  each of the input parameters are shown in Fig. 4,  
whereas  those  which display a confused picture were omitted.
The diagrams  indicate that \Haa increases  with the shock velocity and with the ionization parameter,
but decreases with the SB effective temperature between 10$^4$K and 10$^5$ K because 
the H$^+$ recombination coefficient increases for  gas  heated by the source radiation
to decreasing temperatures  $<$ 10$^5$K.
The SB temperatures  in the hosts  of the LGRB sample  presented in Table 2 were found  within the norm  
10$^4$ $<$\Ts $<$ 10$^5$ K and  the shock velocities  are \Vs$\geq$ 100 \kms.  
The maximum temperature of the gas  downstream is therefore  $\geq$ 1.5 $\times$10$^5$ K (see Sect. 2)
leading  to energies $\geq$ 13 eV,  similar to the  H ionization potential (13.6 eV). 
Shocks with \Vs$>$100 \kms  enhance the  SFR throughout the host galaxies.
Moreover, \Haa increases with $U$, the ionization parameter,  when more photons reach
the emitting nebula  leading to strong  line emission. 
The low ionization parameters $U$  calculated for the Berg et al  and Marino et al samples (Contini 2017)
yield a low \Haa. However, we will show in the following that the SFR  are different for the two samples.

Before  discussing the SFR  trend obtained by the observations we check the \Ha/\Hb ratios presented by the
observers after correction. 
  In our  previous investigation  of SFR in LGRB host galaxies (Contini 2016, fig. 8)  
 \Ha/\Hb=3 (Osterbrock 1974) was used to correct the spectra from reddening because  it is
  adapted  to represent the physical conditions throughout the nebulae  when shocks are at work.
 The  line fluxes collected from the observations  were reddening corrected by the observers
 using Cardelli et al (1989) method. Marino et al and Berg et al used \Ha/\Hb=2.86.
Fig. 5 shows that there is an increasing trend of \Ha/\Hb (presented by the observers) with z, although most 
of the values are in the appropriate range (3$\pm$0.5 Osterbrock 1974).
 This indicates that the correction is not complete for some objects and/or that  matter  with a higher optical depth
 may contribute to the spectra,
 yielding  higher \Ha/\Hb  line ratios (Osterbrock 1974) in particular in the redshift range between approximately
 0.25 and 2.5. A valid correction of line fluxes may yield different values of SFRs.

 \begin{figure}
\centering
\includegraphics[width=9.5cm]{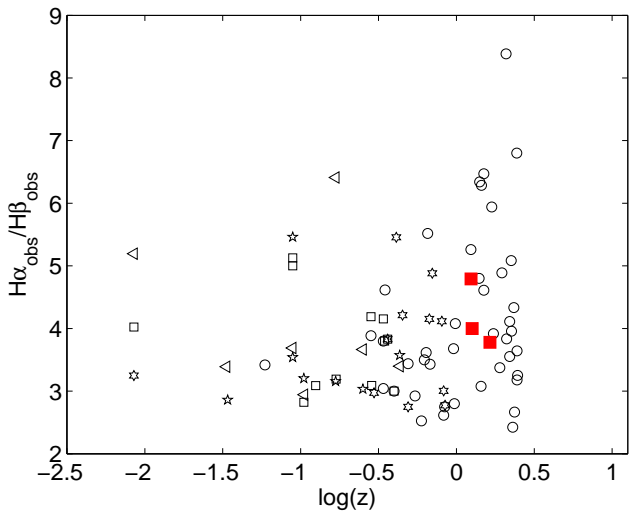}
\caption{Observed \Ha/\Hb  as a function of z.
Black open circles: Kruhler et al.; black pentagrams: Han et al.; black
 open squares: Niino et al.;
black open triangles: Savaglio et al.;
black hexagrams: LGRB (Contini 2016, table8);  red filled squares: Palmerio et al.
}
\end{figure}

\begin{figure*}
	\centering
\includegraphics[width=8.8cm]{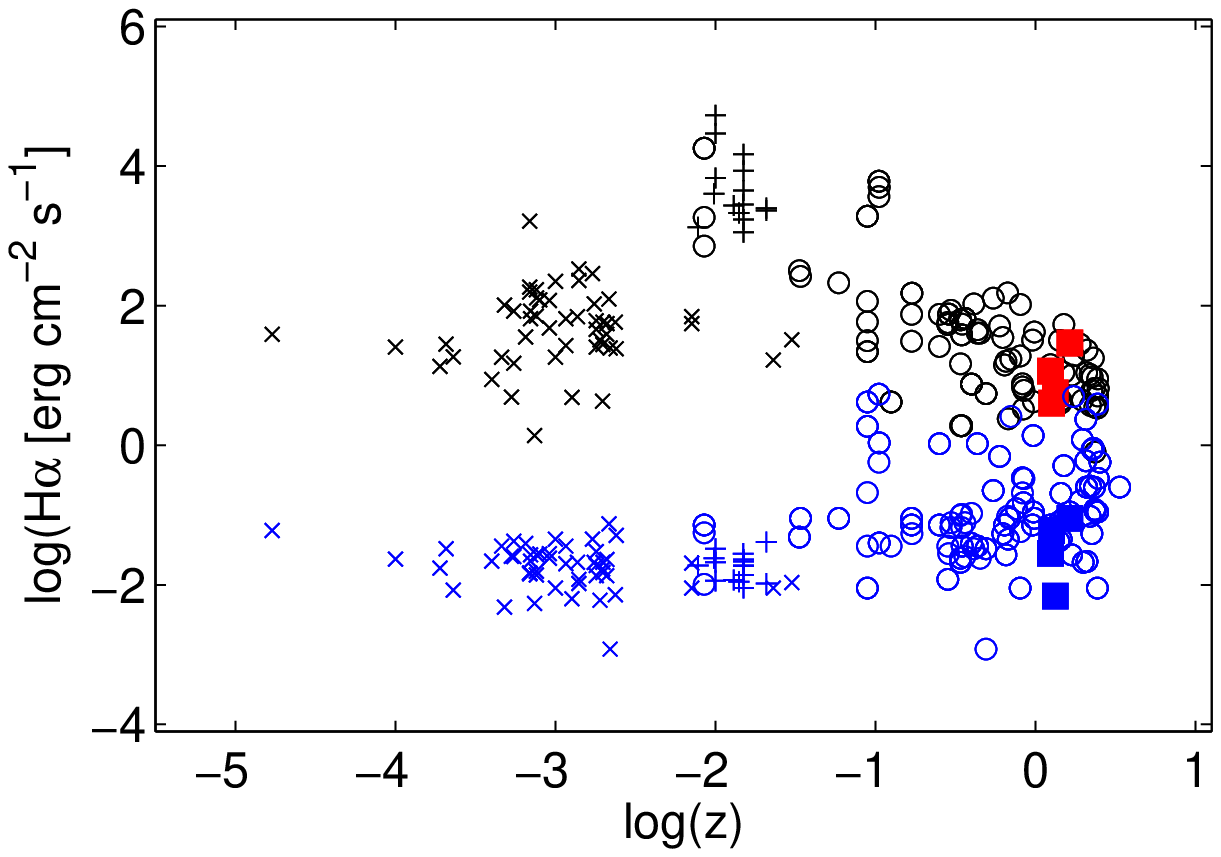}
\includegraphics[width=8.8cm]{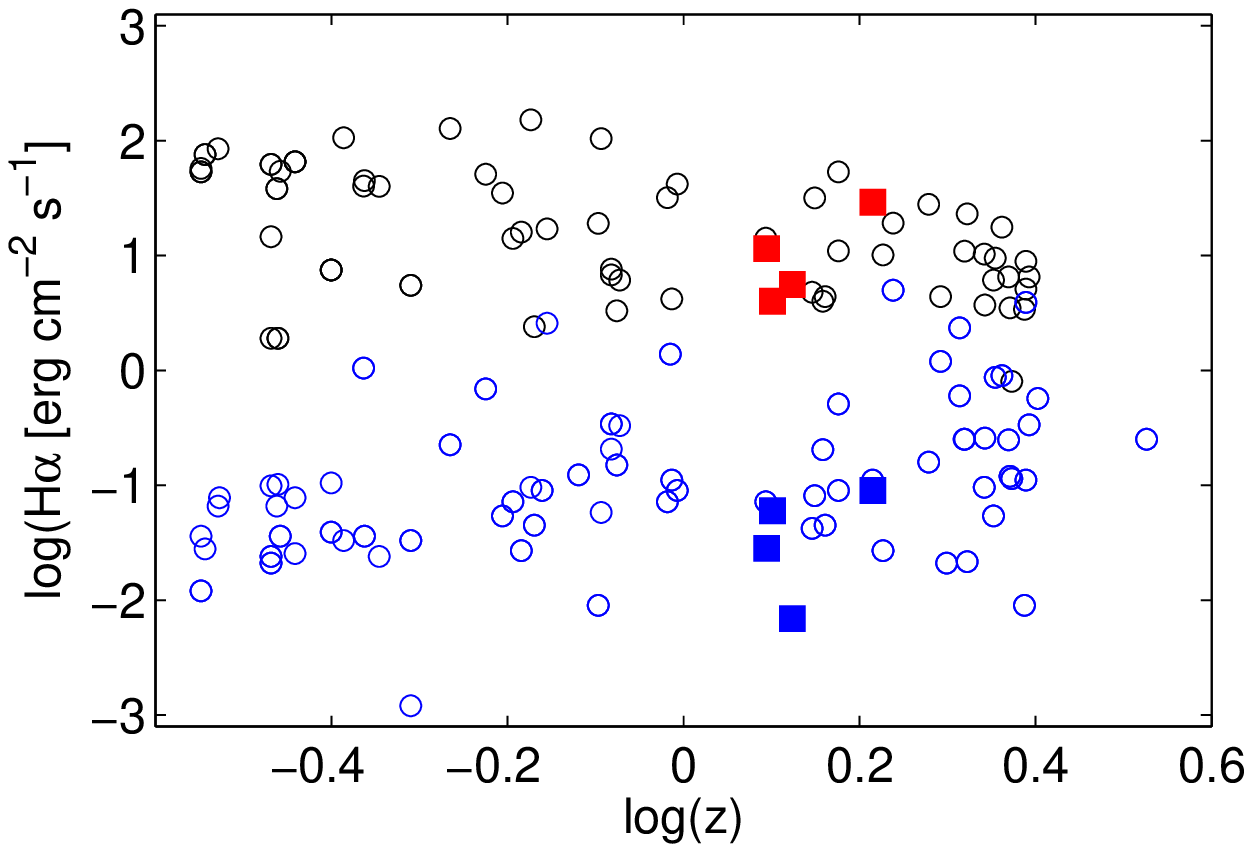}
	\caption{Comparison of  log(\Hab) measured by the observers (black symbols)  with log(\Haa) calculated by 
	the detailed modelling of the spectra (blue symbols).	The galaxies are  described in Table 3.
	Log(\Hab) for LGRB (open circles), log(\Hab) for HII regions in local complexes (plus),
	log(\Hab) for nearby low-luminous HII galaxies (cross).
	Red filled squares: \Hab from the Palmerio et al sample. Blue filled squares: \Haa calculated on the basis
	of the Palmerio et al data.
	}

	\centering
\includegraphics[width=8.8cm]{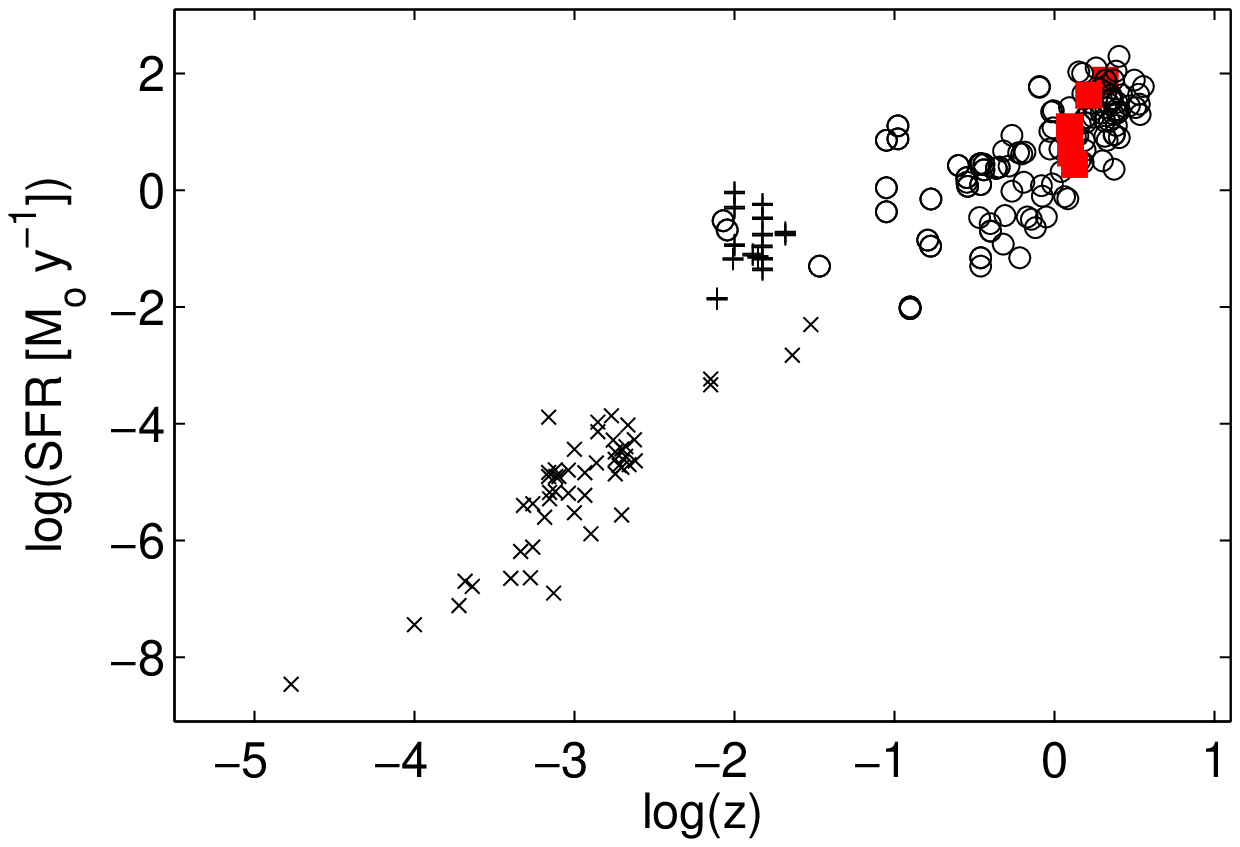}
\includegraphics[width=8.8cm]{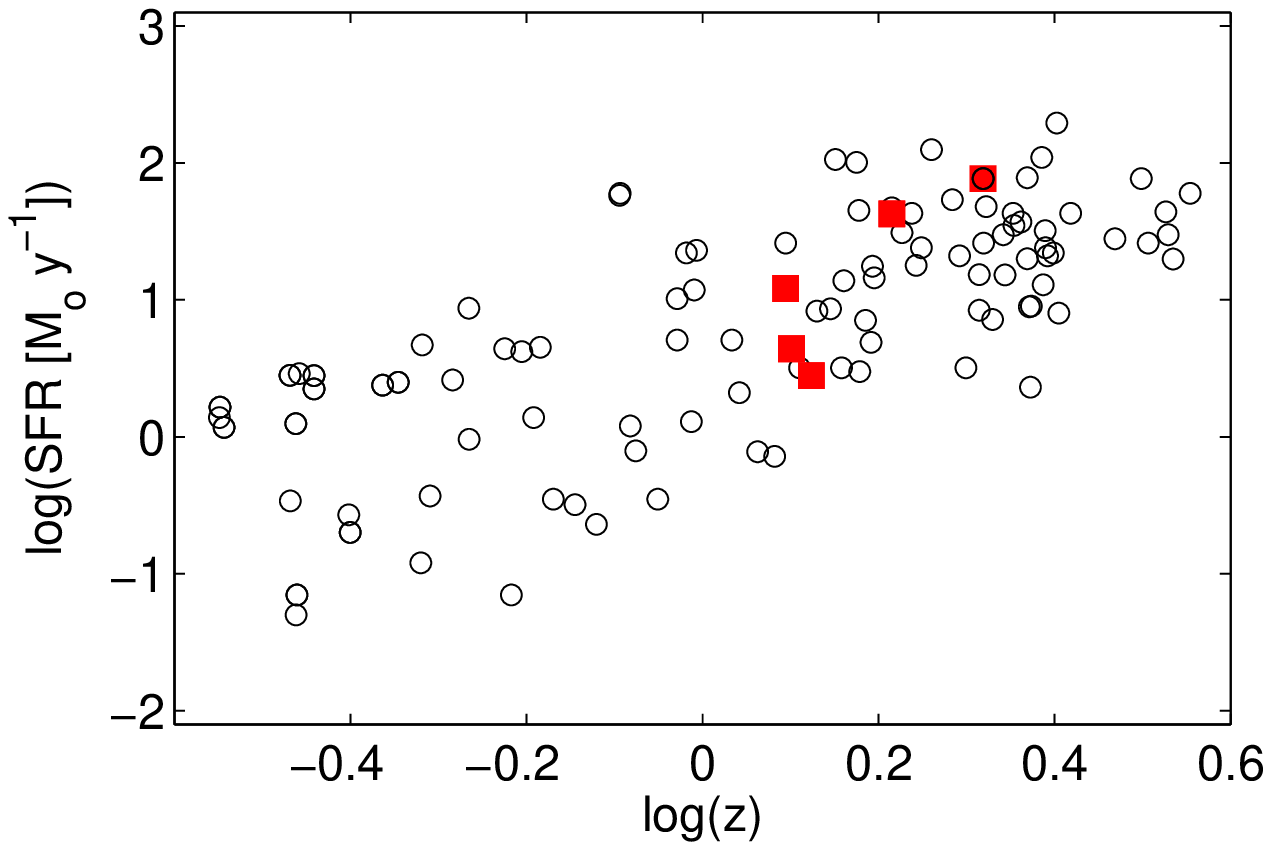}
        \caption{Distribution of log(SFR) along the redshift presented by the observers.
        Red filled squares: Palmerio et al (2019);
        black plus: Marino et al. (2013);
        black crosses: Berg et al (2012); black open circles:
        Kr\"{u}hler et al. (2015),
        Savaglio et al. (2009),
        Han et al. (2010),
        Sollerman et al (2005), Castro-Tirado et al (2001), Graham \& Fruechter (2013),
        Levesque et al (2010), Vergani et al (2011), Piranomonte et al (2015); Niino et al. (2016).
}

\centering
\includegraphics[width=8.8cm]{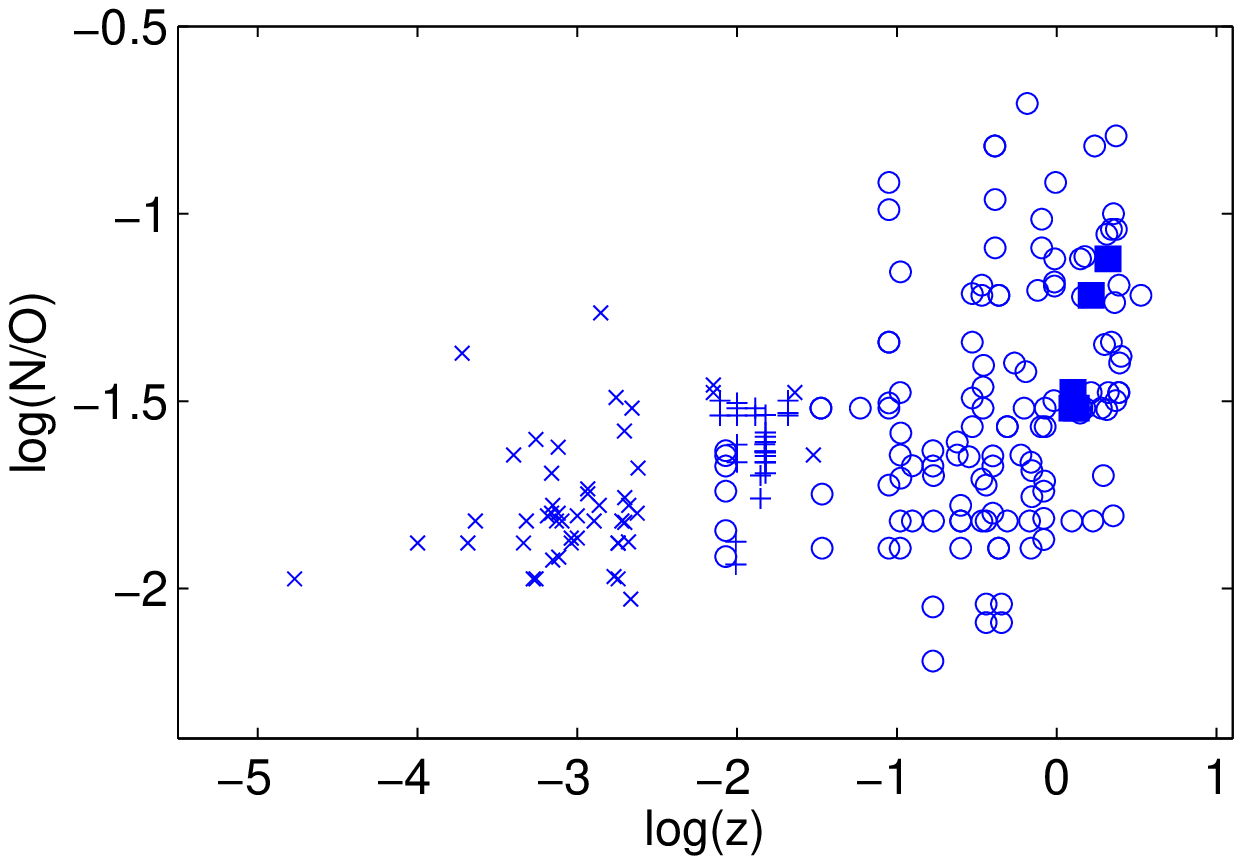}
\includegraphics[width=8.8cm]{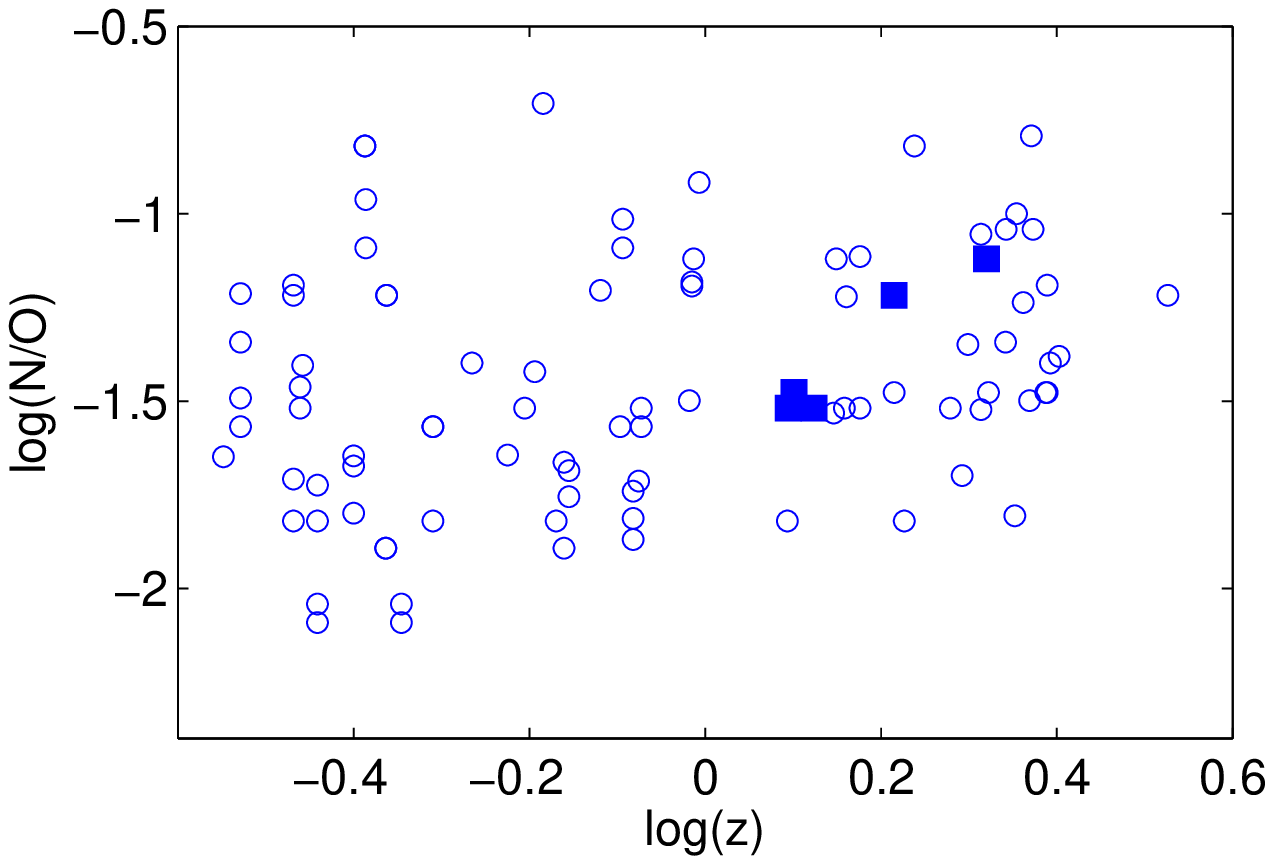}
	\caption{Distribution of calculated log(N/O) abundance ratios  along  the redshift.
Symbols as in Fig. 6.
}                                                                                                    
\end{figure*}

\subsection{SFR}

We compare in Fig. 6 \Haa with \Hab line intensities. \Hab fluxes are shifted upwards on the Y-axis by a factor 
of 10$^{17}$.  The left panels of Figs. 6 - 8  cover a large redshift  range
-5.5$ < $log(z)$ < $0.6 while  the right ones are zoomed to  -0.6$ < $log(z) $ <$0.6. \Haa and \Hab  follow  different trends as
 \Hab  decreases with z, while  the \Haa trend increases, in particular towards higher z.
This means that also the SFR trend should   decrease at high z because SFRs are  generally calculated by eq. 1. 
The opposite can be seen for SFR in Fig. 7 (right panel).
 The left panel of Fig. 7
 shows the SFR values  collected from the observer samples
throughout  a large z  range (z$\leq$3). It seems that the distance factor d$^2$ which  is used in 
the calculation of \La dominates grossly the SFR trend  from local HII galaxies to  LGRB hosts.
SFR  could show a steeper increasing trend at high z if the  spectra were corrected  with a higher precision.
 SFRs presented by  Marino et al CALIFA HII regions in near galaxies (Fig. 7 left) overcome by
 a factor of $\leq$ 100  the SFR values calculated by Niino et al for LGRB host at z$<$0.4, 
while the  Berg et al sample of local galaxies follows roughly  a  linear increasing slope.
We suggest that the abnormally high SFR values in the CALIFA HII  regions  are due to the fact 
that  Marino et al consider  HII $regions$ within the   galaxies which 
are  more   compact than  the entire $galaxies$  observed  by  e.g. Berg et al.
Galaxies, in general,  have  strongly  fragmented structures. 
The \Vs and $U$ parameters   calculated for  the Berg et al sample  
of HII galaxies and  for the Marino et al HII regions (Contini 2017)
are alike, therefore the \Haa values are similar and the different luminosities  depend on  the morphological
distribution of matter through the galaxy.
The   abnormal  SFR (and \Hab) in the  Marino et al HII compact regions and other nearby
galaxies at z$\leq$0.1 is most probably  due   
to  a  high filling factor \ff.  The low \ff~ which  is revealed by the 
fragmented  texture of  the galaxy medium  most probably  depends on the  progenitor merging phenomena.

\subsection{N/O abundance ratios}

The formation of stars  depends on  dust production. 
 Table 2 shows that the calculated  N/O ratios are lower than solar, while O/H are solar. 
We have checked whether the release of  O$_2$ and N$_2$  molecules trapped into the   ice mantles of dust grains 
 can explain the low 
N/O relative abundances calculated for the sample of LGRB hosts presented by Palmerio et al.
O$_2$ and N$_2$ molecules  are trapped within  grain ice mantles with an O$_2$ efficiency 
 greater than for  N$_2$ (Laufer et al. 2018). They are released  with  roughly similar ratios by 
  heating the ice  to the critical temperatures of the exothermic morphological transformations, 140K-160K 
  from amorphous to crystalline cubic and 160K-190K from cubic to hexagonal.
Ice mantles are completely destroyed for \Vs$\sim$ 50 \kms (Strahler \& Palla 2005).
The shock velocities calculated by the detailed modelling of the spectra for a relatively large sample of LGRB 
hosts and HII galaxies (Contini 2017) are in general $\geq$100 \kms.
Therefore, a low N abundance in the gaseous phase of LGRB host clouds does not seem  to be due to trapping of N$_2$ 
molecules into ice mantles.

SFR trends are generally discussed  in the light of massive stars and therefore of metallicities
intended as  the O/H relative abundances (e.g. Vincenzo et al 2015, Mannucci et al 2010).
We focus on  N/H ratios.
Nitrogen is mostly a secondary element being a product of CNO cycle formed at expences of C and O already
present in stars. The primary N component originating in giant branch stars (AGB) is predicted by stellar
nucleosynthesis (Chiappini et al 2003). 
Henry et al (2000) have found that N has both primary and secondary components.
The production of
nitrogen is dominated by primary process at low metallicity and secondary process at higher ones 
(Maynet \& Maeder 2002). 
Contini (2017) showed that O/H and N/H ratios have a minimum for 0.1$\leq$z$\leq$0.4, while N/O decreases towards low z.
For z$<$0.1 the O/H ratios are close to solar  within a large redshift range. In the same redshift range
the  N/O ratios follow an increasing (AGN, SN hosts, etc ) or a decreasing (SB , HII galaxies) trend with decreasing  z
for different types of objects. 
 It was suggested by Contini (2017 and references therein) that  N/O ratios  depend
on  secondary nitrogen production.
The N/O ratios reach  the  lowest  values in  local HII galaxies. 
For comparison, in Fig. 8  the
results  of N/O calculated for the host galaxy sample presented by Palmerio et al  are  added in the 
diagrams showing the calculated N/O  versus z trend   through the extended redshift range.
They are too few to  confirm the decreasing trend of N/O with decreasing z  at z$\leq$2, however they
fit  the average  distribution of N/O for z$\sim$1-2 (Fig. 8 right). 
Fig. 8 (right) shows that the N/O  trend decreasing  with z is similar to that  indicated 
by Fig. 7 for  SFR. The CNO cycle which is responsible for secondary N is a very temperature sensitive process.
Reid \& Hawley (2005) claim that a self mainaining CNO chain would require  high core temperatures 
(T$\geq$16$\times$10$^6$K) and masses M$>$1.5\Mo. They are not predicted at z$<$0.1 by Fig. 8 (left).

\section{Concluding remarks}

We have  calculated the physical conditions and the element abundances
of recently observed GRB host galaxies at redshift 1$<$z$<$2.1 by modelling the spectra  presented by 
Palmerio et al (2019) and Hashimoto et al (2018).  We have compared them
with those found for  previously analysed  LGRB hosts by Contini (2017 and references therein) on a large redshift range.
GRB061007, GRB061121, GRB080605 and GRB090926B  spectra were presented by Palmerio et al (2019).
 We have added  the line spectrum  of GRB080207  from the sample of Kr\"{u}hler et al and
 the continuum SED of both GRB080207 and  GRB070521  from
  Hashimoto et al (2018) because they show a rich collection of  photometric  data.
Other samples, such as  e.g. that of Vergani et al (2017) could not be used because  the 
spectra do not contain  enough data to constrain a detailed modelling.
We have found that  the physical conditions throughout the hosts and the element abundances are 
consistent with those found  for  other hosts in  the same redshift range.
We have investigated the distribution of SFR  throughout the extended  redshift range 0.00001$<$z$<$4
adding to the SFR sample  collected for LGRB  hosts  in previous works, 
the  SFR  observed in nearby HII regions by Marino et al (2013) and
in the  low luminosity local HII galaxies by Berg et al (2012). 
SFR are analysed by the \Ha  fluxes. 
It  seems that the log(SFR) trend  is quasi-linear with log(z)  on a large z range.
We suggest that the   SFR   values  presented by  Marino et al for HII regions are higher than
those given by Berg et al for local HII galaxies  at z$\leq$0.1     
because  the   filling factor   is higher in  compact HII  regions  rather  than  throughout entire galaxies.
The merging process  of the progenitors could lead to a highly disomogeneous morphological structure
of the ISM.

Comparing  the SFR trend  with that  of the N/O abundance ratios, we have found that both decrease for z$<$3.
We have checked whether the release of O$_2$ and N$_2$ molecules trapped into the dust grain ice mantles
by  the exothermic morphological 
transformations of  water ice   could  explain the N/O ratios calculated  throughout the redshift.
However, shock velocities $\geq$ 50 \kms completely destroy the  ice mantles.
The  shock velocities calculated by the detailed modelling of the spectra are generally $\geq$ 100 \kms
for LGRB hosts and SB galaxies. 
 Therefore, the  prevention  of   secondary N formation  for z$<$ 1  remains  a valid hypothesis to explain 
 the  decreasing trend of N/O ratios  towards low z.


\end{document}